\documentclass{emulateapj}

\begin{document}

\bibliographystyle{apj}

\title{Stabilizing a Fabry-Perot etalon to 3 cm/s for spectrograph calibration}

\author{C. Schwab\altaffilmark{1,2}, J. St\"{u}rmer\altaffilmark{3}, Y. V. Gurevich\altaffilmark{4}, T. F\"{u}hrer\altaffilmark{5}, S. K. Lamoreaux\altaffilmark{4}, T. Walther\altaffilmark{5}, and A. Quirrenbach\altaffilmark{3}}

\altaffiltext{1}{Department of Astronomy \& Astrophysics, The Pennsylvania State University, University Park, PA 16802}
%\altaffiltext{2}{Center for Exoplanets \& Habitable Worlds, The Pennsylvania State University, University Park, PA 16802}
\altaffiltext{2}{NASA Sagan Fellow}
\altaffiltext{3}{Landessternwarte, ZAH, K\"{o}nigstuhl 12, D-69117 Heidelberg, Germany}
\altaffiltext{4}{Department of Physics, Yale University, P.O. Box 208120, New Haven, CT 06520}
\altaffiltext{5}{TU Darmstadt, Institute of Applied Physics, Schlossgartenstrasse 7, 64289 Darmstadt, Germany}

\submitted{\sc Submitted to PASP}

\begin{abstract}
We present a method of frequency stabilizing a broadband etalon that can serve as a high-precision wavelength calibrator for an Echelle spectrograph. Using a laser to probe the Doppler-free saturated absorption of the rubidium $D_2$ line, we stabilize one etalon transmission peak directly to the rubidium frequency. The rubidium transition is an established frequency standard and has been used to lock lasers to fractional stabilities of $<10^{-12}$~\citep{Affolderbach,Ye1996}, a level of accuracy far exceeding the demands of radial velocity (RV) searches for exoplanets. We describe a simple setup designed specifically for use at an observatory and demonstrate that we can stabilize the etalon peak to a relative precision of $<10^{-10}$; this is equivalent to 3~cm/s RV precision.
\end{abstract}

\keywords{astronomical instrumentation; extrasolar planets}

\section{Introduction}

Detecting an earthlike planet in the habitable zone of a solar-type star requires a radial velocity (RV) precision of a few cm/s, an order of magnitude better than the best currently attainable precision, which is $\sim70$~cm/s, of which 30~cm/s is due to the wavelength calibration uncertainty~\citep{Lovis2006, Dumusque2013}. To detect a velocity change of 3~cm/s, a relative wavelength shift of $v/c=10^{-10}$ must be measured at a typical spectrograph resolution of only $10^5$. Detecting a shift of only $10^{-5}$  of a resolution element is an extraordinary challenge, and the calibration has to track changes in the spectrograph on that level. Currently available calibration technologies using thorium-argon (ThAr) emission lamps or gas absorption cells are not sufficiently precise for this.

With a range of new spectrographs under construction or coming into operation over the next 5-10 years~(ESPRESSO~\citep{Espresso}; CODEX~\citep{Codex}; NRES~\citep{Nres}; IRD~\citep{Ird}; CARMENES~\citep{Quirrenbach2010}; HPF~\citep{Hpf}; HRS~\citep{Hrs}; SPIRou~\citep{Spirou}), there is clearly demand for a reliable and affordable calibration method with intrinsic precision and long-term stability in the cm/s range. Two techniques currently being explored for producing calibration spectra that meet this requirement are laser frequency combs (LFCs) and Fabry-Perot etalons. While LFCs promise to exceed the requirements on precision and stability, they are very complex systems with a high cost of ownership. A broadband Fabry-Perot etalon illuminated with white light produces a spectrum of regularly spaced lines, creating a ``passive comb" that is far simpler and less expensive than a LFC; however, the etalon does not provide an absolute frequency reference and must be stabilized to be useful for calibration. Nevertheless, impressive results have been achieved using passively stabilized etalons for spectrograph calibration~\citep{Wildi2012}.

In this paper, we describe a method of stabilizing a Fabry-Perot etalon and referencing it to a known absolute frequency by locking it to an atomic transition. We use saturated absorption spectroscopy to stabilize one transmission peak of a broadband etalon to a hyperfine transition of rubidium. The rubidium transition is widely used as a frequency standard, which has been utilized to stabilize lasers to relative precisions of $<10^{-12}$~\citep{Ye1996,Affolderbach}; it provides an ideal, repeatable, and stable wavelength reference. Our locking scheme is simple and robust and will work for a wide variety of etalons operating in the visible and near-infrared (NIR).

\section{Wavelength calibrators}

\subsection{Emission lamps}
The classic method of calibrating a spectrograph is to take emission lamp spectra either before and after the astronomical observation or simultaneously with the observation using a second optical fiber. In the visible part of the spectrum, ThAr hollow cathode lamps are commonly used. The thorium lines provide an absolute wavelength solution for the Echelle spectrum. However, the spectral properties of ThAr lamps are far from ideal for calibration - the lines are unevenly distributed in frequency, have very different brightness, and most lines are blended at typical spectrograph resolution~\citep{LovisPepe2007}. The total lifetime of emission lamps is only a few hundred hours~\citep{Murphy2007}, and their spectra change with age~\citep{Mayor2009}. These effects can be mitigated by using a system of master and slave lamps~\citep{Quirrenbach2010}, where the slave lamps are calibrated against the master lamps, which are used very sparingly. This procedure makes the use of ThAr lamps complicated and time consuming. Obtaining high quality ThAr lamps is also becoming more difficult due to increased restrictions on the thorium needed for their production (Hans Dekker and David Buckley, priv. comm.). ThAr lamps are the primary calibration source for the HARPS spectrograph, and~\citet{LovisPepe2007} report a calibration precision of 20~cm/s with the simultaneous calibration method. In the near-infrared (900-1300~nm), uranium-neon emission lamps are used for calibration ~\citep{Ramsey2010}, which suffer from similar problems as ThAr lamps.

\subsection{Gas absorption cells}
The second common method of calibrating a spectrograph is to pass the starlight through a gas cell~\citep{Butler1996,Campbell1979}, which imprints known, stable absorption lines onto the stellar spectrum. This method has the advantage of tracking changes of the slit (or fiber) illumination; any changes in the input illumination to the spectrograph, for example due to guiding errors, affect the calibration and the stellar spectrum in the same way. This allows the use of slit-fed spectrographs for m/s RV observations. The major drawbacks of gas cells are their limited wavelength range (500 to 630 nm is typical for iodine cells, which are most commonly used~\citep{Marcy1992}) and the high signal to noise ratio (SNR) they require. Reaching the same RV precision using the iodine cell technique requires a SNR at least twice as large as that needed for the simultaneous ThAr calibration~\citep{Bouchy2001}. A gas cell illuminated with a broadband light source can be used in much the same way as an emission lamp~\citep{Mahadevan2009,Redman2012}, which could mitigate the SNR problem. Attempts have also been made to extend the wavelength coverage by finding new gases or mixtures of gases that have absorption lines in the NIR, but with moderate success~\citep{Mahadevan2009,Redman2012,Plavchan2013}.

\subsection{Laser frequency combs}
A laser frequency comb consists of a mode-locked laser whose repetition rate is locked to an atomic clock; this produces a comb-like spectrum of narrow lines whose frequencies have a fractional accuracy and long-term stability of $<10^{-12}$~\citep{Li2008}, far better than is needed for RV measurements. The spectral lines of an LFC are very bright compared to an emission lamp, a distinct advantage for reaching excellent SNR and saving exposure time in calibration frames. Tests by~\citet{Wilken2012} validate the impressive precision attainable with LFCs: they report 2.5 cm/s calibration uncertainty when comparing two simultaneously recorded LFC spectra with HARPS.

An important technical challenge of using LFCs for calibration is the line spacing: since the frequency separation between adjacent comb lines must be resolved by the spectrograph, lines must be separated by $>7.5$~GHz, which requires either an extremely high repetition rate for the laser or sending the comb through one or more Fabry-Perot filter cavities to remove most of the lines, which adds technical complexity. Also, the line intensity varies widely across the bandwidth~\citep{Ycas2012}, so that methods to flatten the gain must be employed~\citep{Probst2013}. Despite considerable effort in the past several years to develop dedicated `astro-combs', no routine operation has yet been achieved. LFCs are very complex and costly systems which currently require significant maintenance to operate; as such, they are only a viable option for wavelength calibration at large facilities.

A promising recent development are so-called microcombs, frequency combs that are not based on a mode-locked laser but on optical pumping of a microoptical cavity~\citep{DelHaye2007}. Due to their larger line spacings, microcombs do not require filtering to be useful for spectrograph calibration, eliminating much of the complexity of astro-combs based on regular LFCs. Indeed, the line spacing is larger than the optimum value for calibration of high-resolution spectrographs. Also, the line spacing depends on the pump power, making it necessary to stabilize the power as well as the frequency of the pump laser, for example by locking two comb lines to two different rubidium transitions~\citep{Papp2013}. Further development will show whether microcombs can be tailored to the specific needs of astronomical applications; if so, they could become a cost-effective alternative to regular astro-combs~\citep{Kippenberg2011}.

\subsection{Etalons}
A Fabry-Perot etalon (FPE) is an optical resonator which transmits light at frequencies that are integer multiples of its free spectral range (FSR). The comb-like spectrum produced by a FPE illuminated with a broadband light source can be used to calibrate a spectrograph. In addition to the FSR, an etalon is characterized by its linewidth; the ratio FSR/linewidth is called the finesse.

\citet{Wildi2010,Wildi2011,Wildi2012} developed a passively stabilized, plane mirror FPE calibrator for HARPS which achieves a nightly wavelength calibration precision of better than 10 cm/s. The etalon is now offered as a calibration mode on HARPS~\citep{Harps} and HARPS-N~\citep{Cosentino2012}. An important consideration for plane mirror etalons is their sensitivity to coupling alignment. For a plane FPE, shifts of the illuminating fiber by a few microns can cause spectral shifts of m/s~\citep{Schaefer2012}. Such shifts could go unnoticed unless the etalon lines are regularly compared with spectral lines of known absolute frequency; on HARPS, this problem is addressed by recalibrating the etalon against a ThAr lamp every night~\citep{Harps}.

\citet{Halverson2012,Halverson2014} are developing a calibrator for the NIR based on a fiber Fabry-Perot (FFP). This etalon consists of a short piece of single-mode fiber (SMF) with mirror coatings on the ends. Light is coupled into and out of the etalon via single-mode fiber pigtails. In this scheme, the light never leaves the waveguide, making FFPs largely insensitive to the coupling conditions. \citet{Halverson2014} demonstrated short term stability of around 80~cm/s with this setup, using the APOGEE spectrograph. This stability is dominated by spectrograph systematics and photon noise, implying that the intrinsic stability of the etalon is likely better.

For both calibration etalons developed to date, the etalon cavity is stabilized by actively controlling the temperature. The sensitivity of the spectral shift to temperature variations is very different for air/vacuum gap etalons compared to solid etalons like FFPs. The tuning rate for a fused silica FFP is a factor of 293 higher than for an air-gap etalon with a Zerodur spacer. To reach 10~cm/s precision, a Zerodur etalon requires 15~mK stability, while a FFP has to be stabilized to 53~$\mu$K. Measuring and controlling the temperature at this level is experimentally challenging~\citep{ILX38}.

\section{Laser-locked etalon}

We take a more direct approach to producing a stable etalon spectrum. Instead of measuring and controlling a secondary indicator like the temperature, we measure the frequency of one etalon fringe directly by referencing it to a hyperfine transition of rubidium (Rb). By probing both the etalon and the Rb simultaneously with the same laser, their frequencies can be directly compared and the etalon peak locked to the Rb peak. This technique places much less stringent requirements on the passive stability of the setup, as the etalon must only be stabilized to within one FSR to ensure that we always lock the same etalon peak. Since the Rb transition is an absolute frequency standard, the etalon can be locked repeatedly and reliably to the same frequency. To control the etalon frequency, we can either change the mirror spacing mechanically (by incorporating a piezoelectric transducer (PZT) into the cavity) or use the temperature of the etalon, which changes the length of the spacer and also the refractive index for a solid etalon. For an air-spaced etalon, tuning by varying the air pressure is also possible. Care must be taken to ensure that no non-common path effects can lead to a shift of the white-light etalon lines compared to the one measured by the laser. In practice, this can be achieved by modal filtering of both the laser and the white light source by feeding them through the same SMF.

We chose Rb as our atomic frequency standard because the Rb $D_2$ transition provides a large absorption cross section, enabling good SNR with a compact gas cell. At 780~nm, the wavelength of this transition is to the red of the bandwidth of typical visible spectrographs and to the blue of the bandwidth of NIR spectrographs; this eliminates contamination of stellar spectra by laser light. Furthermore, laser diodes and optical components at 780~nm are readily available. Other elements such as sodium, potassium, cesium, lithium, or iodine could also be used if their transition frequencies are better suited for a particular study.

Our locking technique provides substantial simplification compared to other methods typically used for stabilizing lasers to atomic transitions, for example lock-in techniques~\citep{Wallard1972} or the Pound-Drever-Hall method~\citep{pdh83,pdhIntro}. Indeed, since the laser is not used as a frequency reference for locking the etalon, but only as a light source for simultaneously recording Rb and etalon spectra, we do not need to lock the laser to the Rb transition at all. A PC-based measurement system and a software proportional-integral-derivative (PID) loop are used to lock the peak position of the etalon directly to the Rb frequency. The advantage of this method is its simplicity: to our knowledge, this is the only scheme that does not require locking the laser to the Rb transition and then locking the etalon to the laser.  While more sophisticated setups provide higher precision, with stabilities of $<10^{-12}$~\citep{Ye1996,Affolderbach}, we adapted our method to the precision requirements of RV studies. Removing the added complexity and hardware necessary for other stabilization methods maximizes reliability and makes for a cost-effective solution that is easy to set up, making our instrument well suited for an observatory environment.

Our method is useful even if a particular calibration etalon has no control parameter with which it can be locked to the Rb frequency or the time needed for the lock to stabilize is too long. In this case, our technique can be used to continuously measure the position of the etalon relative to the Rb frequency, making it possible to correct RV measurements for etalon drift, even without locking the etalon. This is similar to the currently used technique of regularly recalibrating the etalon against an emission lamp, but our method provides real-time drift correction and much more precise knowledge of the etalon's frequency offset.

\section{Apparatus}

\subsection{Laser}

\begin{figure}[tb]
\begin{center}
\includegraphics[width=\columnwidth]{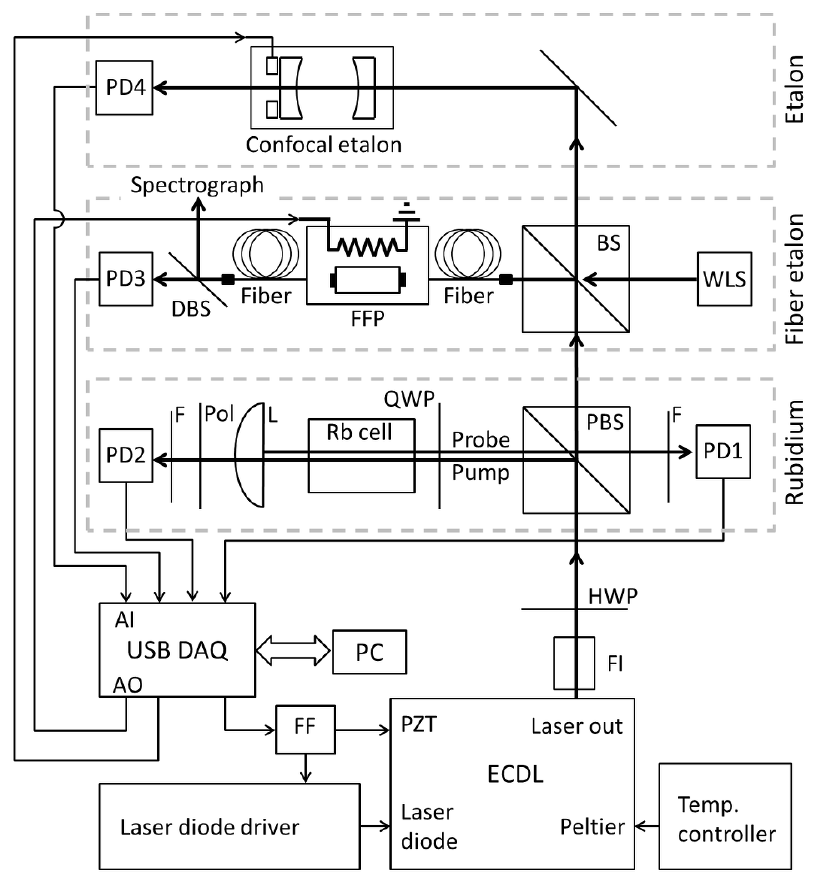}
\caption{Diagram of experimental setup. ECDL: external cavity diode laser, FF: feedforward, FI: Farady isolator, HWP: half-wave plate, PD: photodiode, F: filter, BS: beamsplitter, PBS: polarizing beam splitter, QWP: quarter-wave plate, L: lens, Pol: polarizer, WLS: white light source, FFP: fiber Fabry-Perot etalon, DBS: dichroic beam splitter, DAQ: data acquisition device.}
\label{apparatus}
\end{center}
\end{figure}

Fig.~\ref{apparatus} shows a diagram of the apparatus. We use an extended cavity diode laser (ECDL) in the Littrow configuration~\citep{ECDLbook,WiemanHollberg1991}, which we built in-house. Our design is similar to~\citet{Ricci1995}. The extended cavity is formed between a diffraction grating (Thorlabs GH13-18V) and the back facet of the laser diode (QDLaser QLF073D), and can be tuned by rotating the grating, in our case using a fine-pitched screw for coarse tuning and a PZT for fine tuning. Two Faraday isolators (Isowave I-780-LM, isolation $>36$~dB)  protect the laser from unwanted optical feedback.

The temperature of the laser diode and grating mount assembly is measured with a thermistor and regulated with a Peltier element using a PID controller (based on Analog Technologies TEC5V6A-D). We cool the assembly to bring the free-running wavelength of the laser diode close to 780~nm. The temperature control loop is enabled at all times to avoid thermal cycling of the setup. We drive the laser using a current source developed at the TU Darmstadt~\citep{Fuehrer3,LibbrechtHall,Durfee}. This laser driver provides extremely low noise and narrow linewidth, typically below 10~kHz~\citep{Fuehrer3}. We did not measure the linewidth of our laser, as the line is not resolved by either a high-finesse scanning FPE or the Rb lines.

The locking technique requires a laser that can scan over all components of the Rb $D_2$ line without mode hops. A tuning range of $\sim$8 GHz will cover all hyperfine lines for $^{85}$Rb and $^{87}$Rb; in practice, only one of the hyperfine groups is needed, so a scan range of $\sim$0.5 GHz is adequate. The tuning range must also be large enough to cover one full peak of the etalon and have a sufficient capture range for the peak. Depending on the stability and linewidth of the etalon, this may require a scan range of up to a few GHz. A mode-hop free (MHF) tuning range of 1-2 GHz is fairly typical for laser diodes without antireflection coating on the output facet, and our laser has a MHF scan range of $\sim3$~GHz. This range can be increased by tuning the diode gain curve and internal cavity (i.e. adjusting the current) simultaneously with the tuning of the grating~\citep{Dutta2012}. Implementing this simple feedforward technique brought the MHF scan range of our laser to 15 GHz, which is certainly sufficient for this application.

Even larger MHF tuning ranges are possible using an active ECDL stabilization technique~\citep{Fuehrer1, Fuehrer2}. One advantage of having a large tuning range for the laser is that it allows us to dither the etalon, that is, move the peak in controlled steps through one FSR to successively illuminate all the pixels along the spectrum. This is useful for mapping out the pixel grid of CCD or IR array detectors, for example to detect discontinuities in pixel size~\citep{Wilken2010}. Another advantage is that a scan range larger than the FSR of the etalon allows us to measure the transmission function of the etalon at very high resolution ($>10^9$).

\subsection{Rubidium spectroscopy}

We use the absorption spectrum of rubidium atoms in a gas cell as our frequency standard. The precision with which the center of the absorption line can be determined is limited by the Doppler width (519~MHz at room temperature). For this reason, we use saturated absorption spectroscopy~\citep{Demtroeder} to remove Doppler broadening from the measurement. This velocity-selective technique can provide absorption features comparable in width to the natural linewidth of the transition, which is 6~MHz for the Rb $D_2$ line~\citep{Steck87}.

\begin{figure}[htb]
\begin{center}
\includegraphics[width=\columnwidth]{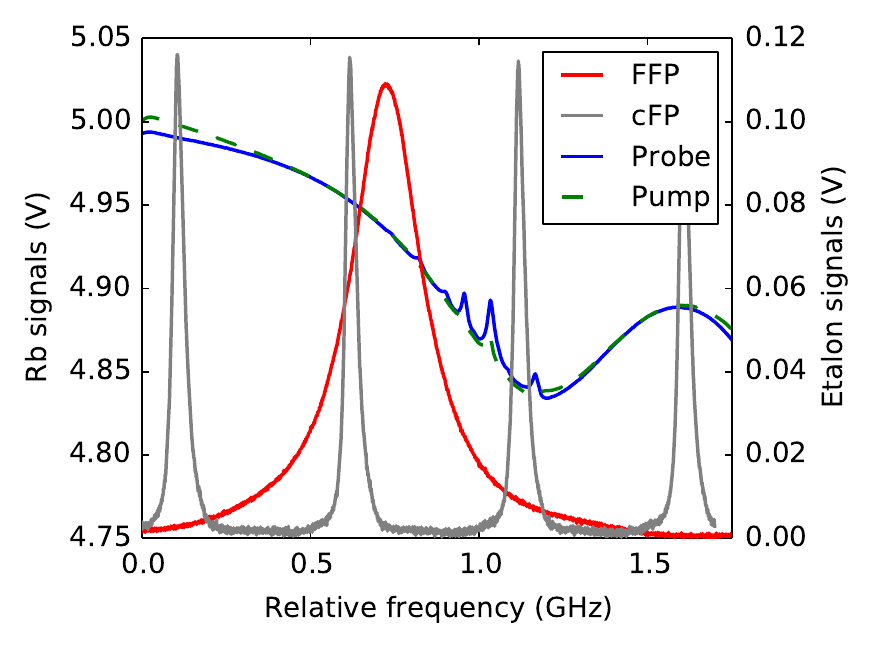}
\caption{Rubidium and etalon spectra. Here we plot a $1.75$~GHz wide scan that spans 88~ms ($\sim3000$ samples). The Doppler-broadened peak of $^{87}$Rb can be seen at 1.2~GHz. The Doppler width is $\sim520$~MHz, and the observed width of the Doppler-free hyperfine lines is 17~MHz. The narrow transmission peaks (36~MHz FWHM) are from a 150~mm confocal FPE ($\mathrm{FSR}=500$~MHz) that we use to monitor the laser. The broad transmission peak (FWHM 250~MHz) is from a silver coated FFP with a FSR of $\sim16$~GHz.}
\label{satAbs}
\end{center}
\end{figure}

We probe the Rb transition with a $1.4$~mm diameter laser beam. The optical power entering the Rb setup is controlled by a half-wave plate and polarizing beamsplitter. After the beamsplitter, the beam passes through a quarter wave plate into the Rb cell; this is the pump beam. An uncoated plano-convex lens (f=25.4mm) is located on the opposite side of the cell, with its planar side facing the cell. 4\% of the incident light is reflected at the plane surface back towards the Rb cell, forming the probe beam. The lens is held in a kinematic mount and adjusted to make the pump and probe beams overlap precisely within the cell. The convex side of the lens does not produce a ghost beam. The Rb cell is tilted by a few degrees to prevent reflections at the cell windows from entering the probe beam photodiode. A second, identical photodiode is used to record the pump beam, which is attenuated by a polarizer to have the same intensity as the probe beam at the photodiode. We operate the system in a temperature stabilized lab and do not regulate the Rb cell temperature. For pump beam intensities comparable to or larger than the saturation intensity ($\sim2$~mW/$\mathrm{cm}^2$ for the Rb $D_2$ line~\citep{Steck87}), saturation broadening increases the width of the Doppler-free features. We use relatively high pump power, which was adjusted to optimize the SNR of the line position measurement and minimize the scatter in the error signal. Fig.~\ref{satAbs} shows typical spectra obtained with our setup.

\subsection{Etalon}

An etalon designed for spectrograph calibration must have a sufficiently large FSR for the separate transmission peaks to be resolved by the spectrograph and should exhibit a linewidth narrower than the spectrograph's line spread function; also, it must produce peaks over the entire wavelength range of the spectrograph. As a working baseline, we assume a minimum FSR of 7.5~GHz (approximately 2 resolution elements on the detector of a high-resolution spectrograph) and a linewidth smaller than 500~MHz (or 0.5 detector pixels).

For locking precision, the linewidth is the only relevant parameter. The position of narrower lines is easier to measure; however, achieving a narrow linewidth requires stricter tolerances for the cavity. An etalon with narrower lines will produce weaker peaks when illuminated with a white light source since the total throughput per line is lower. Supercontinuum sources or laser-driven light sources generate sufficient spectral flux for this to not be an issue. If the etalon lines are sufficiently narrow, they can be approximated as delta functions, and their image on the detector provides a direct measurement of the spectrograph's line spread function.

\begin{figure}[tb]
\begin{center}
\includegraphics[width=\columnwidth]{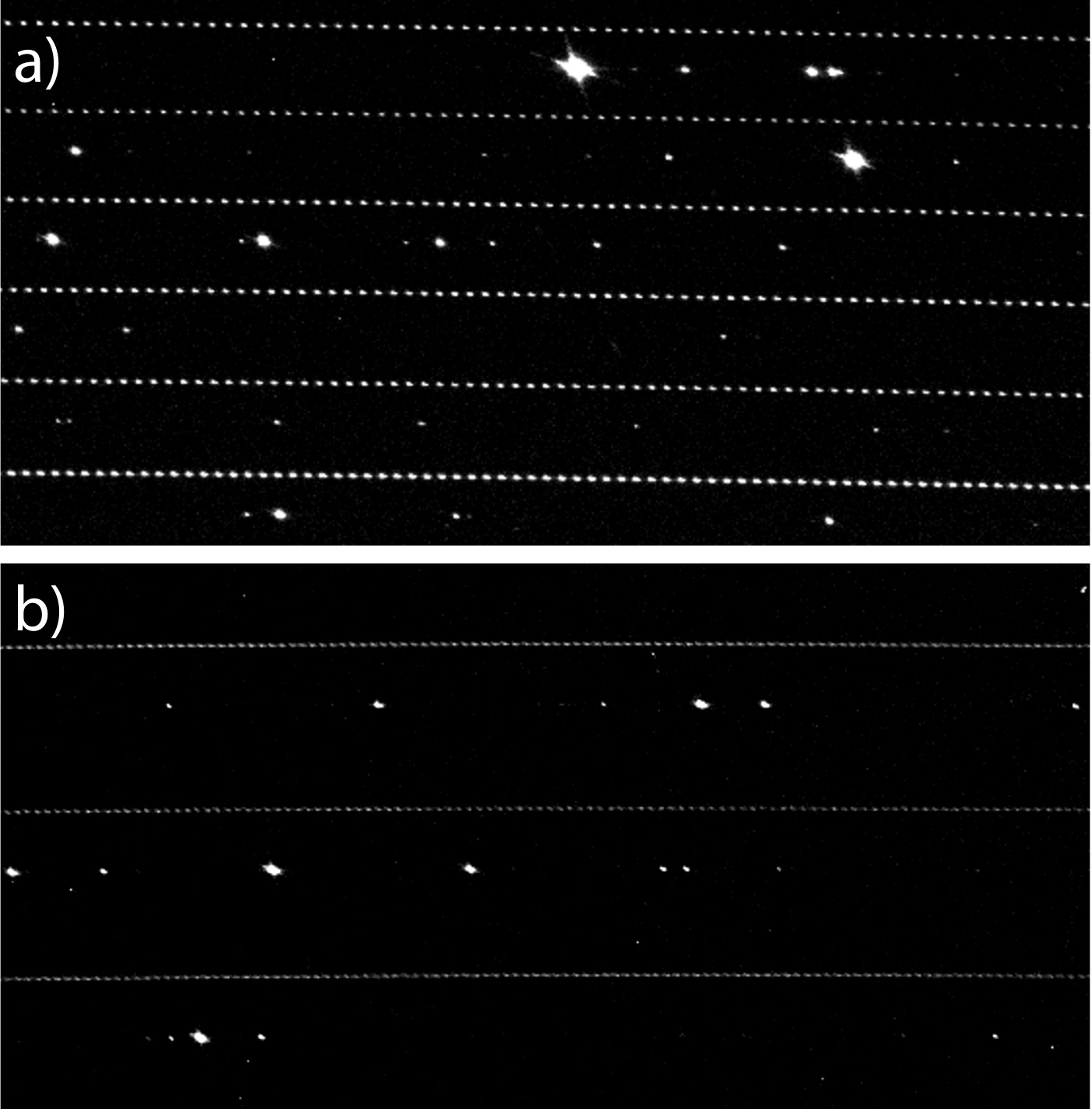}
\caption{Spectra of a ThAr lamp and the FFP illuminated by a white light LED. The spectra were taken consecutively using our laboratory Echelle spectrograph, and the ThAr frame was shifted in software by 30 pixels towards the blue in the cross-dispersion direction. \textbf{a)} The section shown here is at the red end of the LED bandwidth. \textbf{b)} A different part of the spectrum, at the blue end of the LED bandwidth.}
\label{spectra}
\end{center}
\end{figure}

Spectrograph calibration requires an etalon that works over a very broad wavelength range. Since custom dielectric coatings with flat reflectivity over the necessary bandwidth are costly, we used a metallic coating instead. Metallic coatings are simple, inexpensive, and have the advantage of a very small dependence of the reflection phase shift on the wavelength. This effect is called group velocity dispersion (GVD), and causes the etalon line spacing to change as a function of wavelength, a highly undesirable feature in a calibration etalon~\citep{Wildi2010, BruceCiddor1960}.
For the measurements discussed here, we used a fiber etalon with a FSR of 16~GHz, well suited for spectrograph calibration. We fabricated the etalon in-house by gluing 780~nm SMF into a standard 2.5~mm ceramic ferrule and polishing it to $\sim6$~mm length. We then applied a partially-reflective protected silver coating to the ends of the ferrule. The metallic coating provides a very wide bandwidth; we see a deeply modulated comb spectrum with clearly distinguished lines over the full wavelength range of our laboratory Echelle spectrograph (410 to 800~nm), as shown in Fig.~\ref{spectra}. The ferrule is inserted into an aluminum fiber mating sleeve, which is in turn held in a tight-fitting copper piece. A resistive cartridge heater (Thorlabs HT15W) is used to heat the assembly and the temperature is measured with a thermistor. We use SMFs to couple light into and out of the FFP. This etalon is similar to the commercial units investigated by~\citet{Halverson2012}.

In addition, we use a 150~mm long confocal etalon ($\mathrm{FSR}=500$~MHz) to monitor the tuning behavior of the laser. This etalon consists of two dielectric mirrors with 97\% reflectivity from 630 to 1120~nm (Layertec 103283) and a thermally compensated spacer made from fused silica and brass~\citep{Barry2013}. A cylindrical PZT (Noliac NAC2124) in the cavity allows the etalon length to be scanned. We have also successfully locked this etalon to the Rb transition with our setup, using the PZT to adjust the cavity length. Although the small FSR of this etalon makes it unsuitable for spectrograph calibration, our ability to stabilize it to similar precision as the FFP demonstrates the generality of our locking technique.

\subsection{Data acquisition and software}

The laser frequency is swept by applying a smoothed sawtooth voltage to the PZT. A National Instruments USB-4431 generates the sawtooth and simultaneously samples the Rb probe, Rb pump, and etalon transmission signals. The voltage is amplified by a factor of $\sim4$ in the feedforward circuit before being sent to the PZT, which allows a maximum scan range of 6.3~GHz without using a high-voltage amplifier.

The lock is controlled using custom software written in Python, using the Qt framework for a GUI. The software subtracts the Rb pump signal from the probe signal to eliminate the Doppler background, leaving only the saturated absorption peaks. For locking, we use the five largest of the six peaks of the $^{87}$Rb $F=2 \; \rightarrow \; F'=1,2,3$ transitions\footnotemark[1]\footnotetext[1]{Three of the peaks correspond to real transitions, while the other three are crossover peaks, which occur at frequencies exactly halfway between each pair of real peaks.}. The precisely known frequency spacing between these peaks~\citep{Steck87} allows us to calibrate our voltage scale in terms of frequency, and we use the average of the five peak locations as the ``reference frequency" to which we lock the etalon. Using the average of multiple peaks instead of the location of a single peak as the lock point allows for better rejection of electrical noise. To find the etalon peak, we fit a Lorentzian to the etalon transmission signal.

The error signal is the difference between the etalon peak position and the Rb reference frequency. A PID control loop implemented in software generates the control signal, which is output by a National Instruments myDAQ card. We test our lock with both the PZT-driven confocal etalon and the thermally controlled FFP. For the PZT-driven etalon, the error signal measured at each loop iteration is sent directly to the PID. For the FFP, due to its slower response, we smooth the error signal using a moving average over the last 200 measurements before sending it to the PID. For both etalons, the control signal is updated at every loop iteration, which corresponds to an update rate of 15 Hz. The myDAQ card output is sufficient to drive the PZT of our confocal etalon directly. To drive the cartridge heater for the FFP, we feed the myDAQ signal into the modulation input of a laser diode driver (Analog Technologies ATLS100mA103-D), which we use as very stable, low noise current amplifier.

\section{Results}

\begin{figure}[tb]
\begin{center}
\includegraphics[width=\columnwidth]{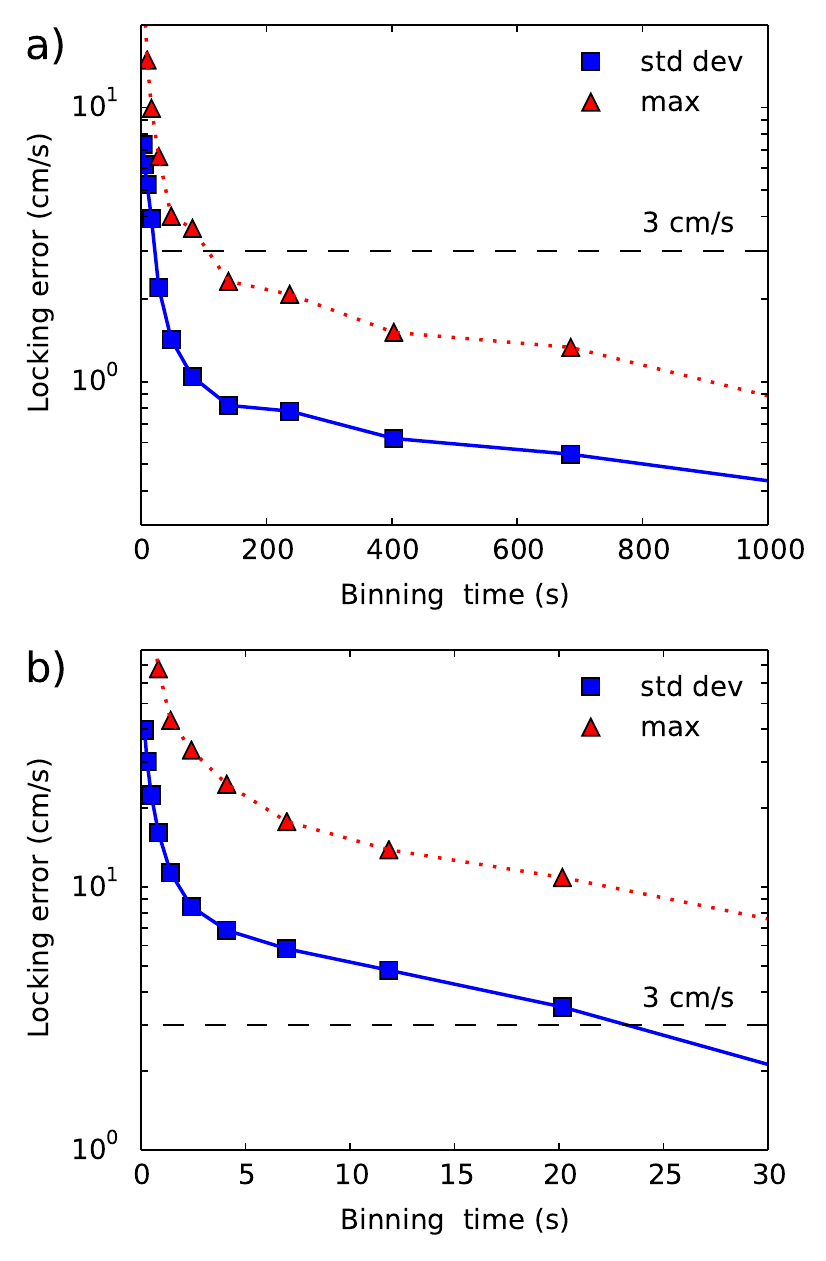}
\caption{Maximum (triangles) and standard deviation (squares) of the locked error signal as a function of bin width (integration time) for the PZT-driven confocal etalon. The dashed line is at 3~cm/s. Panel b) shows the first 30~s of panel a).}
\label{piezoLock}
\end{center}
\end{figure}

The bandwidth of the PZT and myDAQ card is high enough to move the confocal etalon peak by one FSR within one cycle of our loop, enabling us to lock this etalon without any further time lag. We stabilized the etalon to the highest peak in the Rb spectrum, determining the Rb peak position simply by choosing the sample with the highest amplitude within the peak. When the lock is running, the standard deviation of the error signal is about 45~cm/s, due largely to the scatter in the Rb peak determination itself. However, the error is normally distributed and drops quickly with integration time, as Fig.~\ref{piezoLock} shows. For an integration time of 23~s, the etalon peak is locked to better than 3~cm/s even without further optimization. For binning times $>15$ minutes, the standard deviation levels out at approximately 4.5~mm/s or a relative accuracy of $1.5 \times 10^{-11}$. Due to the bandwidth of the lock, optimizing the PID parameters is straightforward, and the lock recovers quickly from short perturbations.

\begin{figure}[tb]
\begin{center}
\includegraphics[width=\columnwidth]{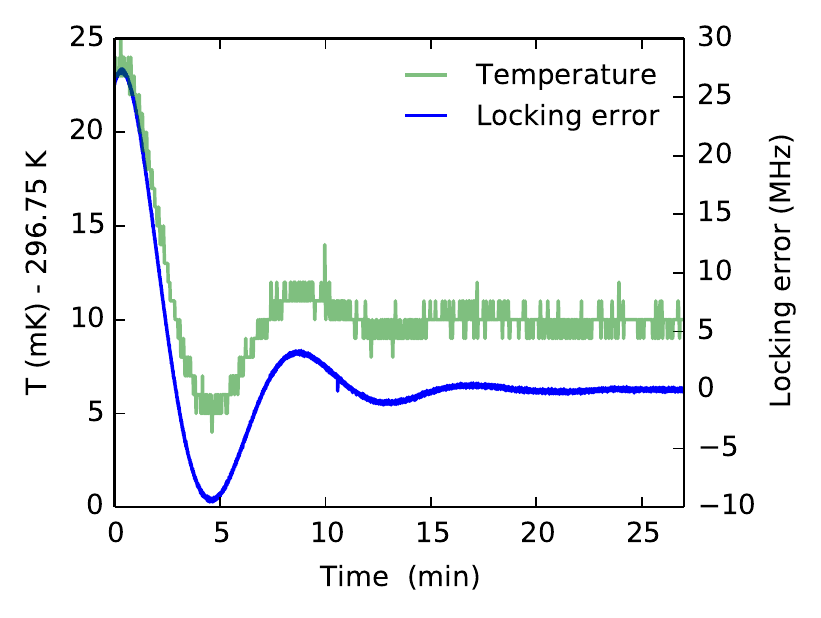}
\caption{Locking error of the FFP and measured temperature of the copper block when starting the PID loop. We measured the temperature with a Pt100 sensor, which was mounted on top of the copper block holding the FFP and read out by a Lakeshore 336 temperature controller. The resolution of this combination is only $\sim1$~mK, but this measurement illustrates the thermal behavior of the copper block. There is no obvious time lag between the temperature probe and the etalon peak frequency on these timescales.}
\label{lockStart}
\end{center}
\end{figure}

\begin{figure}[tb]
\begin{center}
\includegraphics[width=\columnwidth]{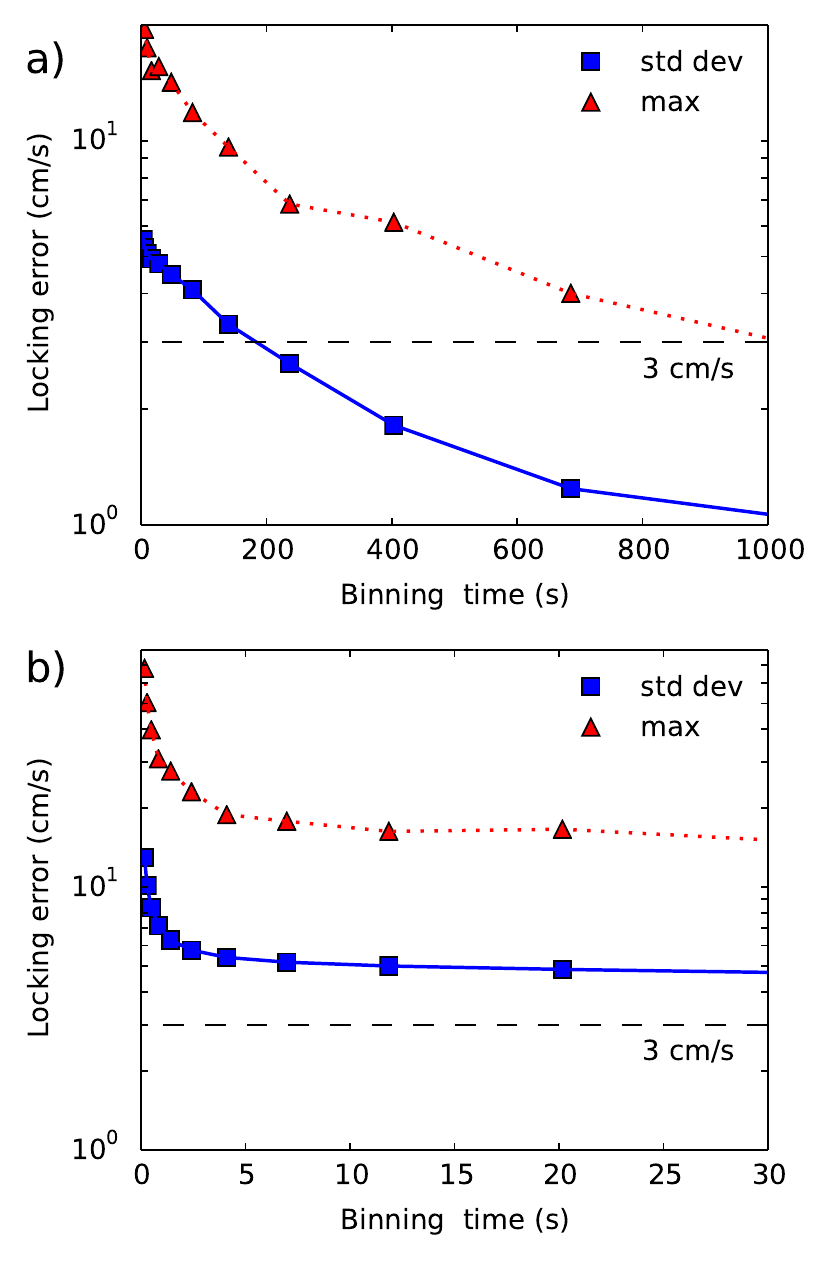}
\caption{Maximum (triangles) and standard deviation (squares) of the locked error signal as a function of bin width (integration time) for the FFP. The dashed line is at 3~cm/s. Panel b) shows the first 30~s of panel a).}
\label{FFPLock}
\end{center}
\end{figure}

For the thermally tuned FFP, the reaction time is significantly longer than for the PZT-driven etalon (Fig~\ref{lockStart}). To accommodate this, we average over many measurements, as described above, and lock the etalon to the Rb reference frequency, which is more stable than the individual peak positions. With the lock enabled, the FFP peak is stabilized with a standard deviation of $<18$ cm/s. The scatter in the data is well approximated by a normal distribution. Fig.~\ref{FFPLock} shows the standard deviation of our measurement of the locking error as a function of bin width. For a bin width corresponding to a typical exposure time of 3~minutes, the standard deviation is 3~cm/s. After integration times $\ge20$~minutes, the standard deviation is $<1$~cm/s. Importantly, our knowledge of the etalon offset from the Rb line, described by the standard error of the mean, is significantly better than the locking precision. For integration times $>20$~s, the standard error of the mean is $<1$~cm/s, so even very short exposures, like calibration frames, can be corrected at that level of precision.

Under normal laboratory conditions, the FFP routinely stayed locked over the longest timescales investigated ($\sim24$~hours). The mean of the error signal in our longest datasets is consistent with zero, showing that the etalon peak reliably tracks the Rb reference frequency. Our setup was located in a temperature stabilized lab with $\pm1$~K temperature variation. The air pressure was not stabilized or monitored. This approximates a typical environment for a fiber-fed high precision spectrograph housed in a Coud\'{e} room. It is important to note that the etalon was not stabilized in any other way, only thermally insulated inside a Styrofoam housing, and the optical table on which the setup is mounted was not vibration isolated. The excellent precision both laser-locked etalons deliver under these conditions demonstrates the power of our technique.

\section{Discussion and outlook}

\subsection{Reliability}
Reliability, the necessary amount of maintenance, and cost of operation are important considerations for subsystems of a spectrograph installed at an observatory. With this in mind, we chose commercial off-the-shelf components for parts that could break. The laser diode is probably the most likely component to fail. However, a mean time between failures of up to 100,000 hours, or $>10$ years of continuous operation, is reported~\citep{ILX33}. We run the laser diode at 50\% of the rated maximum current and cool it below ambient temperature, both of which improve lifetime. Similarly, the PZT that scans the laser is used far below its maximum voltage rating. Instead of biasing the PZT, we use a fine-pitched screw to adjust the frequency of the laser to the Rb line. Since we scan fairly slowly, the power dissipation inside the PZT is low. Lastly, we smooth out the sawtooth applied to the PZT to avoid stress at the turning points. Changing the laser diode or PZT is straightforward. Maintenance of the laser head does not introduce any offset when the lock is resumed, since the frequency zero point is derived directly from the Rb transition.

Several external parameters can affect the observed Rb transition frequency in a saturation spectroscopy setup like ours and must be carefully controlled. The most important of these are pump beam intensity, angle between the pump and probe beams, Rb cell temperature, and ambient magnetic field. \citet{Affolderbach} measured the line shifts due to these parameters using a setup similar to ours and demonstrated sufficient control of all of them to achieve $< 2 \times 10^{-12}$ relative frequency stability of an ECDL locked to the Rb $D_2$ line over timescales $\ge10^4$~s. Controlling all above-mentioned shifts sufficiently well to guarantee stability and reproducibility of the Rb frequency at the level of precision needed for calibration is relatively straightforward, for example by surrounding the Rb cell with a single layer of mu-metal and actively stabilizing the cell temperature and pump laser power.

The most critical part of the setup is the etalon itself. A replacement etalon will have a slightly different line spacing and probably a different dispersion. It will need to be calibrated against an absolute frequency standard that covers the entire bandwidth, ideally a laser frequency comb, and this calibration compared to the one of the original etalon. However, breaking the etalon assembly seems unlikely. For long term stability, a thermally adjusted etalon seems to be a more robust choice, as in this case the cavity is completely passive, while breaking the PZT in a PZT-driven etalon would render the etalon unusable.

Finally, metallic mirror coatings on the etalons could change over time. The protected silver coatings we employ can degrade through corrosion. For bulk etalons operated inside a vacuum housing, this is unlikely. Noncorroding gold coatings can be used for etalons in the NIR. The coating of an FFP experiences mechanical stress since the cavity fiber is in physical contact with the input and output fibers. Also, depending on the power of the white light source and laser, the flux at the mirrored ends of the single-mode waveguide that forms the etalon can be very high, potentially damaging the coating.  Very dense, low loss dielectric mirror coatings are likely better equipped to withstand these conditions.

\subsection{Future work}

Current work is focused on integrating the system into a compact and rugged unit that can be used at an observatory. This includes improved shielding against electrical interference, improved thermal isolation of the setup, and enclosing the Rb cell in a magnetic shield. To reduce the laser's sensitivity to mechanical vibrations and thermal fluctuations, we plan to implement a simple yet effective active stabilization scheme based on polarization spectroscopy~\citep{Fuehrer2}. We will also optimize the algorithm we use to determine the Rb reference frequency, which we believe is a major contribution to our error budget.

We plan to record simultaneous ThAr and laser-locked FPE spectra with our $R=80,000$ laboratory Echelle spectrograph to map out the dispersion of the etalon and determine the wavelength of each peak. The precision of this measurement will be limited by the precision of the ThAr calibration. The stability of the entire etalon spectrum can be verified with a precision equal to that of our locking technique by simultaneously measuring a second etalon line using a second laser and another atomic transition, for instance the cesium line at 852~nm. Another possibility is frequency doubling a 1560~nm laser and using the second harmonic at 780~nm in the same way we use our laser~\citep{Masuda2007}, but also using the original wavelength to monitor an etalon line at 1560~nm, which would work well for NIR etalons. For etalons in the visible, the same thing can be accomplished by frequency doubling a 780~nm laser and monitoring an etalon line at 390~nm in addition to the original 780~nm line used for locking.

\subsection{Summary}
In summary, we have presented a simple setup that reliably locks an etalon suitable for calibrating an Echelle spectrograph with better than 3~cm/s radial velocity precision for any realistic exposure time. Care was taken to adapt the locking scheme to the specific requirements of an observatory. We use a minimum of optical and electronics hardware and use a single locking loop that references the etalon directly to the rubidium transition. This technique can be used to stabilize a variety of etalons in the visible and NIR, and we have demonstrated excellent precision with thermal as well as PZT-driven locking of air-spaced and fiber-based etalons. Our setup combines long lifetime and high reliability with low cost. It can be easily integrated with any of the etalons currently investigated for astronomical use. Even if a particular etalon cannot be locked to the rubidium transition (i.e. due to a very long time constant), our scheme allows precise real-time monitoring of the etalon drift relative to the rubidium. In this way, observations calibrated using the etalon can be accurately corrected for etalon drift. A laser-locked etalon provides high precision, long-term stability, and a wide spectral range for calibration, making it an excellent, cost-effective solution for calibrating an Echelle spectrograph.

\section*{Acknowledgements}
We thank Sidney Cahn and Stephen Irons for help with the Rb spectroscopy setup and David DeMille for useful discussions about lasers. We thank Andrew Szymkowiak and Ulrich Gr\"{o}tzinger for advice and practical help with the electronics. We acknowledge Lutz Geuer for making the mechanical parts and Julien Spronck for useful discussions at the beginning of this work. We are grateful to Suvrath Mahadevan for his support and advice, and to him and Samuel Halverson for sharing their insight into FFPs. Thanks to Christoph Koke for help with the software, Christopher Tillinghast for the silver coatings, and Karin Loch for help with the data analysis. CS acknowledges Ivelina Momcheva, Rosalind Skelton, and Erik Tollerud for their assistance in completing this project. This work was performed in part under contract with the California Institute of Technology (Caltech)/Jet Propulsion Laboratory (JPL) funded by NASA through the Sagan Fellowship Program executed by the NASA Exoplanet Science Institute.

\bibliography{etalonRefs.bib}

\end{document}